\documentstyle[prl,aps,epsf,amstex,multicol,times,bbm]{revtex}

\def\QED{\leavevmode\unskip\penalty9999 \hbox{}\nobreak\hfill
     \quad\hbox{\leavevmode  \hbox to.77778em{%
               \hfil\vrule   \vbox to.675em%
               {\hrule width.6em\vfil\hrule}\vrule\hfil}}
     \par\vskip3pt}
\def\qed{\leavevmode\unskip\penalty9999 \hbox{}\nobreak\hfill
     \quad\hbox{\leavevmode  \hbox to.77778em{%
               \hfil\vrule   \vbox to.675em%
               {\hrule width.6em\vfil\hrule}\vrule\hfil}}
     \par\vskip3pt}

\def\ibb #1{\leavevmode\hbox{\kern.3em\vrule
     height 1.5ex depth -.1ex width .4pt\kern-.3em\rm#1}}

\newtheorem{prop}{Proposition}

\newcommand{\proofend}{\hfill\rule[0pt]{.2cm}{.2cm} }

\begin{document}


\title{The entangling power of passive optical elements}

\author{Michael M.\ Wolf$^{1}$, Jens\ Eisert$^{2}$, and Martin B.\ Plenio$^{2}$}

\address{1 Institute for Mathematical Physics, TU Braunschweig,
38106 Braunschweig, Germany}

\address{2
QOLS, Blackett Laboratory, Imperial College of Science, Technology
and Medicine,
London, SW7 2BW, UK}

\date{\today}

\maketitle

\begin{abstract}
We investigate the entangling capability of passive optical
elements, both qualitatively and quantitatively. We present a
general necessary and sufficient condition for the possibility of
creating distillable entanglement in an arbitrary multi-mode
Gaussian state with the help of passive optical elements, thereby
establishing a general connection between squeezing and the
entanglement that is attainable by non-squeezing operations.
Special attention is devoted to general two-mode Gaussian states,
for which we provide the optimal entangling procedure, present an
explicit formula for the attainable degree of entanglement
measured in terms of the logarithmic negativity, and discuss
several practically important special cases.
\end{abstract}


\begin{multicols}{2}
Entangled states of light field modes may be generated by
transmitting two squeezed states through a beam splitter
\cite{Theo}. This is one of the experimentally accessible
procedures for generating continuous-variable entanglement in
optical systems \cite{Exp}. Moreover, it is a particular example
of a situation where passive optical elements exhibit their
entangling power when applied to Gaussian input states. It is well
known that the presence of squeezing is necessary for obtaining
entanglement in this manner \cite{Theo}. However, the degree of
the attained entanglement is by no means the same for all input
states: it depends to a large extent on the degree and direction
of squeezing of the incoming modes and on the specific properties
of the beam splitter. This raises the question under what
circumstances such an entangling procedure is optimal in the sense
of generating states which have the maximal attainable amount of
entanglement. And in general, by means of arbitrary passive
optical elements, what are the requirements such that entanglement
can be generated between any bi-partite split of a system in a
multi-mode Gaussian state?

In this letter we address the question of the entangling power of
passive optical elements acting on any number of modes in an
arbitrary Gaussian state, qualitatively as well as quantitatively.
Passive optical operations can be implemented by using beam
splitters and phase shifters \cite{ReckZeil}. These are cheap
operations and easy to implement in contrast to squeezing
operations. Therefore we will consider squeezing as a potential
resource for entanglement and ask for the requirements and the
optimal way of entangling a squeezed state by means of passive
operations, which we assume to be available in arbitrary
quantities.
The main result and starting point is a necessary and sufficient
condition for the possibility of creating distillable entanglement
on general Gaussian initial states -- pure or mixed --  between
any bi-partite split of an $n$-mode system with the help of
passive optical elements. We then introduce a lower bound for the
attainable degree of entanglement measured in terms of the
logarithmic negativity \cite{Neg} for $n$-mode systems. Moreover,
we derive a general formula for the largest degree of entanglement
of an arbitrary subsystem consisting of two modes. The operations
that can be implemented with passive optical elements can be
identified with the non-squeezing operations. In this sense we
establish a quantitative connection between the degree of
squeezing of a Gaussian state and the degree of entanglement that
is attainable with the application of non-squeezing operations. Of
particular interest is the case where only two modes are present.
We will discuss  this situation in more detail by explicitly
constructing the optimal entangling procedure and discussing
several meaningful special cases.

\begin{figure}
\centerline{
    \epsfxsize=6.5 cm
       \epsfbox{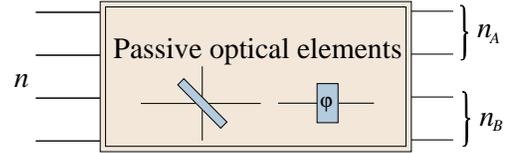}
}
\vspace*{0.2cm}
 \caption{ $n$ field modes in a Gaussian state
$\rho$ can be (NPPT)-entangled with respect to a partition into
$n_A+n_B=n$ modes by means of passive optical elements if and only
if $\lambda_1\lambda_2<1$, where $\lambda_1$ and $\lambda_2$ are
the two smallest eigenvalues of the covariance matrix associated
with $\rho$.}
\end{figure}

We start by introducing the formalism that we will use
extensively. {\em Gaussian states} are completely characterized by
their first and second moments, where only the latter, given in
terms of a covariance matrix $\Gamma$, carry information about
entanglement and squeezing. For this reason we will set the first
moments to zero, which can always be achieved by unitary
operations on individual modes. The {\em covariance matrix} is
then given by $\Gamma_{kl}=2 \langle R_k R_l\rangle-i
\sigma_{kl}$, $k,l=1,\ldots,2n$, where the vector
$R=(Q_1,\ldots,Q_n,P_1,\ldots,P_n)$ consists of the canonical
coordinates for $n$ modes and the {\em symplectic matrix}
\begin{equation}\label{sigma}
\sigma=\left(\begin{array}{cc} 0 & {\mathbbm{1}}_n\\-{\mathbbm{1}}_n & 0
\end{array}\right)
\end{equation}
governs the canonical commutation relations (CCR)
$[R_k,R_l]=i\sigma_{kl}$. A matrix represents an admissible covariance matrix
if it satisfies the Heisenberg uncertainty
relations $\Gamma + i \sigma\geq 0$.
{\em Symplectic transformations}
$\Gamma\longmapsto S^T\Gamma S$  preserve the CCR and
therefore  satisfy $S^T\sigma S=\sigma$ \cite{SMD}. All
symplectic transformations correspond to unitary {\em Gaussian
operations} \cite{CPM} on the level of states, in the sense that
the Gaussian character of arbitrary input states is preserved
under such unitary operations.
They can be decomposed \cite{SMD,BlochMessiah} into {\em
active/nonlinear} and {\em passive/linear} operations
\cite{linear}. The latter can be implemented by using passive
optical elements such as beam splitters and phase plates only
\cite{ReckZeil}, and are of the form $\Gamma\longmapsto K^T\Gamma
K$,
\begin{equation}\label{K}
K = \Omega^\dagger\left(\begin{array}{cc} U & 0\\0 & \overline{U}
  \end{array}\right)\Omega = \left(\begin{array}{cc} X & Y\\-Y &
  X \end{array}\right).
\end{equation}
Here $U=X+iY$ ($X,Y$ real) is any unitary matrix and
\begin{equation}\label{Omega} \Omega =
{1\over\sqrt{2}}\left(\begin{array}{cc} {\mathbbm{1}}_n& i {\mathbbm{1}}_n\\
{\mathbbm{1}}_n & - i {\mathbbm{1}}_n
  \end{array}\right)
\end{equation}
relates real and complex representations by mapping
creation/annihilation operators to position/momentum operators via
$\Omega (R_1,\ldots,R_{2n})^T=(a_1,\ldots, a_n, a_1^\dagger,\ldots,
a_n^\dagger)^T $. Transformations of the type $\Gamma\longmapsto
K^{T}\Gamma K$ with $K$ as above will from now on be denoted as
{\em passive transformations}. Any such $K$ is both symplectic and
orthogonal, i.e., $K^TK={\mathbbm{1}}$, and the set of all
symplectic transformations that can be implemented with passive
optical elements form a group, the maximal compact subgroup $K(n)$
of the group of symplectic transformations $Sp(2n,{\mathbbm{R}})$
\cite{SMD}.

A Gaussian state is said to be {\em squeezed}
if there exists a basis in
phase space such that at least one
diagonal element of the covariance matrix is smaller than one.
From now on we order the eigenvalues of $\Gamma$ in non-increasing
order, so that this implies that the smallest eigenvalue
$\lambda_1$ of $\Gamma$ is smaller than one \cite{SMD}. Since
every passive transformation $K$ is orthogonal, it does not affect
the squeezing of a state.

Let us now turn to entanglement properties. A Gaussian state of a
bi-partite system consisting of parts $A$ and $B$ with $n_A+n_B=n$
modes is {\em separable}, i.e. unentangled between $A$  and $B$,
iff there exist covariance matrices $\Gamma_A,\Gamma_B$ for $n_A$
resp.\ $n_B$ modes such that $\Gamma\geq \Gamma_A\oplus\Gamma_B$
\cite{WW,ABsplit}. A necessary and for $1\times n_B$ modes also
sufficient condition for separability \cite{WW,CiracSimon} is that
the {\em partial transpose} of the state is positive
semi-definite. This, in turn is equivalent to $\Gamma\geq
i\tilde{\sigma}$, with the partially transposed symplectic matrix
$\tilde{\sigma}=({\mathbbm{1}}_n\oplus E)\sigma
({\mathbbm{1}}_n\oplus E)$  and
$E={\mathbbm{1}}_{n_A}\oplus(-{\mathbbm{1}}_{n_B})$ being the
partial transposition operator that reverses all momenta on one
side. It has been shown that a Gaussian state is distillable,
i.e., that its entanglement can be revealed using
    local operations and classical communication,
    iff its partial transpose
    is non-positive \cite{distill}.

Obviously, every entangled Gaussian state is squeezed since
$\lambda_1\geq 1$ would mean that $\Gamma \geq
{\mathbbm{1}}\oplus{\mathbbm{1}}$ which in turn implies
separability. Hence, a state can only be entangled by means of
passive operations if it is squeezed initially. The following
proposition gives a necessary and sufficient condition for the
possibility of transforming a general Gaussian state into a
distillable one by means of passive transformations (see Fig.\ 1):

%
%
%

\begin{prop}\label{main}
Let $\Gamma$ be a covariance matrix corresponding to a Gaussian
state of $n$ modes. A passive transformation $\Gamma\longmapsto
\Gamma'=K^T\Gamma K$ leading to an entangled state having a
non-positive partial transpose with respect to a
 partition into
$n_A+n_B=n$ modes exists iff
\begin{equation}\label{l1l2}
\lambda_1 \lambda_2 < 1,
\end{equation}
where $\lambda_1, \lambda_2$ are the two smallest eigenvalues of
$\Gamma$.
\end{prop}

{\it Proof:}
A Gaussian state of an $n$-mode system with
covariance matrix $\Gamma'$ has a positive partial transpose iff
all symplectic eigenvalues of the respective partially transposed
covariance matrix $\tilde{\Gamma}'= ({\mathbbm{1}}_n\oplus
E)\Gamma' ({\mathbbm{1}}_n\oplus E)$ are larger than or equal to
one. The symplectic eigenvalues of $\tilde{\Gamma}'$ are in turn
equal to the square roots of the ordinary eigenvalues of
$-(\Gamma'\tilde{\sigma})^2$. The square of the smallest
symplectic eigenvalue additionally minimized over all passive
transformations is thus given by
\begin{equation}\label{nu}
\nu := \inf_K \inf_{||\xi||=1} \langle \xi |\Gamma^{1/2}
M^T\Gamma M \Gamma^{1/2}|\xi\rangle,
\end{equation} where $M:=K\tilde{\sigma}K^T$
and $\|.\|$ denotes the standard vector norm.
Hence, we have to
show that inequality (\ref{l1l2}) is equivalent to $\nu<1$.

Since $M$ is an antisymmetric orthogonal matrix, it maps any real
unit vector onto the
two-dimensional unit sphere of its
orthogonal complement. The  vector
\begin{equation}\label{xiprime}
|\xi'(K,\xi)\rangle:= M\Gamma^{1/2}|\xi\rangle \cdot
\langle\xi|\Gamma|\xi\rangle^{-1/2}
\end{equation}
therefore satisfies
$\langle\xi|\Gamma^{1/2}|\xi'(K,\xi)\rangle=0$. Inserting
Eq.\ (\ref{xiprime}) in Eq.\
(\ref{nu}) we get
\begin{eqnarray}\label{nuprime1}
\nu &=& \inf_K \inf_{||\xi||=1} \langle \xi|\Gamma|\xi\rangle
\langle \xi'(K,\xi)|\Gamma|\xi'(K,\xi)\rangle\\ &\geq&
\inf_{||\xi||,||\xi'||=1} \langle \xi|\Gamma|\xi\rangle
\langle\xi'|\Gamma|\xi'\rangle,\label{nuprime2}
\end{eqnarray}
where the  infimum in Eq.\ (\ref{nuprime2}) is taken over all real
unit vectors satisfying $\langle\xi|\Gamma^{1/2}|\xi'\rangle=0$.
This relaxes the requirement that $|\xi'\rangle$
has to be of the form
in Eq.\ (\ref{xiprime}) and therefore leads to the lower bound.
%
%
The minimum in Eq.\ (\ref{nuprime2}) is now attained for vectors lying
in the two-dimensional space corresponding to the two smallest
eigenvalues $\lambda_1, \lambda_2$ of $\Gamma$. Hence,
$|\xi\rangle = \cos\phi |\lambda_1\rangle + \sin\phi
|\lambda_2\rangle$ for some $\phi$ and $|\xi'\rangle
\propto\sqrt{\lambda_2}\sin\phi |\lambda_1\rangle -
\sqrt{\lambda_1} \cos\phi |\lambda_2\rangle$. However, every
$\phi$ leads to the same value  and we have
\begin{equation}\label{nuineq}
\nu \geq \lambda_1 \lambda_2,
\end{equation}
showing that $\lambda_1\lambda_2<1$ is indeed necessary for
$\nu<1$.

In order to prove sufficiency, we have to show that there always
exists a passive transformation $K$ such that $|\xi'\rangle$
is of the
form (\ref{xiprime}) and the inequalities
(\ref{nuprime2},\ref{nuineq}) thus become equalities.
Note that
this is in turn equivalent to the statement that for every pair of
orthogonal real unit vectors $|\xi\rangle \perp|\eta\rangle$
there is a passive
transformation $K$ such that $\langle\eta|K\tilde{\sigma}
K^T|\xi\rangle=1$.
We first show that the problem can be reduced to a
two-mode problem.
Let $|\lambda_1\rangle$ and
$|\lambda_2\rangle$ be the eigenvectors associated with
$\lambda_1$ and $\lambda_2$.
Decomposing $|\lambda_1\rangle ,|\lambda_2\rangle$
into position and momentum components, one may define the
complex form of $|\lambda_1\rangle$ and
$|\lambda_2\rangle$ according to
\begin{eqnarray}
|\Psi_i\rangle :=|\lambda_i^{(Q)}\rangle +i|\lambda_i^{(P)}\rangle,\,\,\,
i=1,2.
\end{eqnarray}
Then there always exists a unitary $U$ such that the vectors
$U|\Psi_i\rangle$ have only two non-zero components in the first two
entries of the vector. In turn, according to
Eqs.\ (\ref{K},\ref{Omega}) this implies that there exists a
passive transformation $S$
such that $\Gamma':=S^T \Gamma S$ has the property that the
leading principle submatrix of $\Gamma'$ associated with
the first two modes has the same  two smallest
eigenvalues $\lambda_1,\lambda_2$.

Similarly, $|\xi\rangle,|\eta\rangle$ can be decomposed into
position and momentum components, and define $|\Psi\rangle
:=|\xi^{(Q)}\rangle+i|\xi^{(P)}\rangle$ and $|\Phi \rangle:=|
\eta^{(Q)}\rangle+i |\eta^{(P)}\rangle$ such that
$\Omega|\xi\rangle=\Omega|\xi^{(Q)}\oplus\xi^{(P)}\rangle=
\frac{1}{\sqrt{2}}|\Psi\oplus\overline{\Psi}\rangle$ and analogous
for $|\eta\rangle $ and $|\Phi\rangle$. Then
$||\Psi||=||\Phi||=||\xi||=||\eta||=1$ and
\begin{eqnarray}\label{Im} \langle\eta|K\tilde{\sigma}
K^T|\xi\rangle &=& {\text{Im}}
\big[\langle\Phi|U E U^\dagger|\Psi\rangle \big],\\
\langle\eta|\xi\rangle &=&
{\text{Re}}\big[\langle\Phi|\Psi\rangle\big]\equiv 0.
\end{eqnarray}
Without loss of generality we fix
$|\Psi\rangle
= (1,0)^T$,
which can always be achieved by applying an additional unitary.
Then, every two-dimensional unit
vector $|\Phi\rangle$ for which
${\text{Re}}\big[\langle\Phi|\Psi\rangle\big]=0$ is of
the form
\begin{equation}\label{Phi}
|\Phi\rangle = -i \big(\cos(2\gamma), e^{2 i
\alpha}\sin(2\gamma)\big)^T.
\end{equation}
Choosing
\begin{equation}\label{V_0}
U=\left(\begin{array}{cc} e^{-i\alpha} \cos(\gamma) &
-e^{-i\alpha} \sin(\gamma)
\\ e^{i\alpha} \sin(\gamma) & e^{i\alpha} \cos(\gamma)
  \end{array}\right),
\end{equation}
we obtain with $E=\mbox{diag}(1,-1)$
\begin{equation}\label{final} 1={\text{Im}} \big[
\langle\Phi|U E U^\dagger |\Psi \rangle
\big]=\langle\eta | K \tilde{\sigma} K^T
 |\xi\rangle ,\nonumber
 \end{equation}
which completes the proof.
\proofend

Whereas every entangled state is squeezed, Proposition \ref{main}
implies that conversely any squeezed state can be entangled by
using passive optical elements supplemented by a single
additional vacuum mode (empty port of a beam splitter),
because the joint covariance matrix $\Gamma\oplus{\mathbbm{1}}_2$
\cite{ABsplit} then satisfies inequality (\ref{l1l2}).
The optimal entangling procedure  consists then of
two steps: (i) One first applies a
passive transformation $S$
such that the smallest eigenvalue $\lambda_1$ of $\Gamma$
is also the smallest eigenvalue of the $2\times 2$ principal
submatrix of $S^T \Gamma S$ corresponding to the first mode.
(ii) One then applies the optimal entangling procedure on
this mode and the vacuum mode, which will be derived in Proposition 3.

The proof of proposition \ref{main} leads to a lower bound for the
attainable entanglement measured in terms of the {\em logarithmic
negativity} \cite{Neg}. The latter is so far the only calculable
entanglement measure for mixed Gaussian states. For an $n$-mode
Gaussian state $\rho$ it is given by
\begin{eqnarray}\label{Neg}
E_{\cal N}
=-\sum_i \text{min} (0, \log_2(s_i)).
\end{eqnarray}
where the  $s_i$, $i=1,\ldots,n$ are the symplectic eigenvalues of
the partially transposed covariance matrix. Since
$\nu=\lambda_1\lambda_2$ is the square of the smallest symplectic
eigenvalue, we obtain
\begin{equation}\label{Ebound}
E_{\cal N}\geq \max (0,-\log_2({\lambda_1 \lambda_2})/2 )
\end{equation}
for the attainable entanglement, with equality if there is only
one $s_i$ smaller than one.
A particularly transparent situation is now the case where we
consider only the entanglement present in an arbitrary two-mode
subsystem obtained when tracing out the other modes at the end.
\begin{prop}\label{n2} Let $\Gamma\longmapsto\Gamma'=K^T\Gamma K$ be a passive transformation acting on a
Gaussian state of $n\geq 2$ modes with covariance matrix $\Gamma$.
The maximum attainable amount of entanglement obtained for an
arbitrary two-mode subsystem of $\Gamma'$ is then given by
\begin{equation}\label{ENmax}
E_{\cal N} = \max\big[0,-\log_2(\lambda_1 \lambda_2)/2\big],
\end{equation}
where $\lambda_1, \lambda_2$ are the two smallest eigenvalues of
$\Gamma$.
\end{prop}
{\it Proof: } First note that for the case of a two-mode state
only one of the two symplectic eigenvalues $s_1,s_2$ of the
partially transposed covariance matrix $\tilde{\Gamma}_{(2)}'$ can
be smaller than one, since $(s_1
s_2)^2=\det{\tilde{\Gamma}_{(2)}'}=\det{\Gamma_{(2)}'}\geq 1$
\cite{det}.
%
Following the same argument as in
Proposition 1, there exists
always a passive transformation $S$ such that a two-mode
principal submatrix
of the covariance matrix $S^T \Gamma S$ has the two smallest
eigenvalues $\lambda_1$ and $\lambda_2$, which leads to equality
in Eq.\ (\ref{Ebound}).
\proofend

A special instance of Proposition \ref{n2} is the case where the
input already is a two-mode system, i.e. $n=2$. For this case we
will now explicitly construct the optimal entangling procedure.
We will show that it is always sufficient to perform a single
phase rotation in one of the two modes, for example in $A$,
succeeded by a beam splitter operation on both modes. Again, it is
most convenient to employ the complex version of the problem. In
their complex forms, a beam splitter $B(\gamma)$ and a phase shift
$L(\alpha)$ in system $A$ are represented by the matrices
\begin{eqnarray}
    B(\gamma)=
    \left(\begin{array}{cc} \cos\gamma &
-\sin\gamma
\\ \sin\gamma & \cos\gamma
  \end{array}\right),
    \,\,\,
    L(\alpha)=\left(\begin{array}{cc}
    e^{-i\alpha} & 0
\\ 0 & 1
  \end{array}\right),
\end{eqnarray}
where $\gamma\in[0,2\pi)$ determines the transmission coefficient
 of the beam splitter, and $\alpha\in [0,\pi)$ is
the phase difference of the incoming and outgoing fields.
Without loss of generality
the beam splitter itself is assumed to induce no phase
difference.

\begin{prop}  Let $\rho$ be a Gaussian
$1\times 1$-mode state with covariance matrix $\Gamma$. Let
$|\lambda_1\rangle$ and $|\lambda_2\rangle$ be the eigenvectors of
the two smallest eigenvalues  of $\Gamma$, with complex versions
$|\Psi_1\rangle$ and $|\Psi_2\rangle$.
The optimal entangling operation using only passive
 optical elements is
given by a  phase rotation $L(\alpha)$ on mode A, followed by a
beam-splitter $B(\gamma/2)$, such that $\gamma$ and $\alpha$ are
the solutions of
\begin{eqnarray}
\cos(\gamma)&=& \text{\rm Im}[
\langle\Psi_2|\sigma_z|\Psi_1\rangle ],\label{j1}\\ \sin(\alpha)
\sin(\gamma) &=& \text{\rm Im}[
\langle\Psi_2|\sigma_y|\Psi_1\rangle ],\label{j2}\\ \cos(\alpha)
\sin(\gamma)&=& \text{\rm Im}[
\langle\Psi_2|\sigma_x|\Psi_1\rangle ],\label{j3}
\end{eqnarray}
where $\sigma_x$, $\sigma_y$, and $\sigma_z$ are the Pauli
spin matrices.
\end{prop}
{\it Proof:}
In order to find the optimal entangling procedure one has to
identify a unitary $V$ such that
\begin{eqnarray}\label{mustbeone}
    \text{Im}
    \left[\langle \Psi_2|
    V E  V^\dagger |\Psi_1\rangle
    \right]=1,
\end{eqnarray}
and decompose it into a beam splitter and a phase shift.
The most general form for $V E V^\dagger=: F$ is given by
\begin{eqnarray}\label{F}
    F(\gamma,\alpha)=\left(
    \begin{array}{cc}
    \cos(\gamma) &  e^{ -i\alpha } \sin({\gamma})\\
    e^{i\alpha } \sin({\gamma}) &  -\cos(\gamma)\\
    \end{array}
    \right),
\end{eqnarray}
which corresponds to $V=L(\alpha) B(\gamma/2)$. Inserting the
decomposition $F(\gamma,\alpha)= \cos(\gamma)\sigma_z +
\cos(\alpha)\sin(\gamma)\sigma_x+\sin(\alpha)\sin(\gamma)\sigma_y$
into Eq. (\ref{F}) one verifies that values $\alpha,\gamma$ that
satisfy Eqs.\ (\ref{j1}-\ref{j3}) provide a solution of Eq.
(\ref{F}). Moreover, the set of equations (\ref{j1}-\ref{j3})
always has a solution, since the vector of the imaginary parts in
Eqs. (\ref{j1}-\ref{j3}) can be shown to be a unit vector if
$\text{Re}[\langle\Psi_2|\Psi_1\rangle]=0$. \proofend

We will in the following apply this result to some special cases.
The covariance matrix $\Gamma$ of the initial state of the
$1\times 1$-mode system will be written in the block form
\begin{equation}\label{sym}
  \Gamma=\left(\begin{array}{cc}
    A & C \\
    C^T & B
  \end{array}\right),
\end{equation}
where $A$ and $B$ are the reduced covariance matrices
corresponding to mode $A$ and $B$ respectively. Depending on the
form of the $2\times 2$- matrices $A$, $B$, and $C$ several
optimal entangling protocols can be identified:

(i) {\it A product of arbitrary single mode Gaussian states:} If
$C=0$, then  a $50:50$ beam-splitter is required in the optimal
entangling procedure. The phase transformation that is needed will
in general depend on the actual form of $A$ and $B$. In
particular:

(ia) {\it A product of two identical single mode states:} In this
case $A=B$ and $C=0$, and one finds that  $\alpha=\gamma=\pi/2$.
The optimal entangling operation is thus a $50:50$ beam-splitter,
which follows a $\alpha=\pi/2$ phase transformation, as expected.
This is the optimal procedure for uncorrelated identical Gaussian
input states used in several experiments \cite{Exp}.

 (ib) {\it A product of a Gaussian single mode state
 and a coherent or thermal state:} In
  this case where $B= b{\mathbbm{1}}$, $b\geq 1$,
  and $C=0$
the optimal entangling operation is again the application of a
$50:50$ beam-splitter. No phase transformation is required.

 (ii) {\it States with covariance matrix in Simon normal form}
 \cite{CiracSimon}: If $A= a{\mathbbm{1}}, B= b{\mathbbm{1}},
 C=\mbox{diag}(c,d)$,
 then one eigenvector $\Psi_i$ is real and the other is imaginary.
 Hence $\alpha\in [0,\pi]$, whereas the optimal beam splitter is in
 general not balanced.

  (iia) {\it Symmetric states:}
These are states with identical thermal reductions, meaning that
$A=B=a {\mathbbm{1}}$, $a\geq 1$. These states are already
optimally entangled,
 since $E_{\cal N}(\rho)=
\max\big[0,-\log(\lambda_1 \lambda_2)/2 \big] $, and the optimal
entangling procedure is thus the identity operation.

(iib) Special cases of symmetric states are {\it two-mode squeezed
pure Gaussian states} with covariance matrix in Simon normal form,
where in addition, $C$ takes the form $C=\text{diag}(c,-c)$ with
$c= (1-a^2)^{1/2}$.

In this letter we have investigated the entangling capabilities of
passive optical elements in a general setting. We have presented a
necessary and sufficient criterion for the possibility of creating
distillable entanglement in a multi-mode system that has been
prepared in a Gaussian state. The findings reveal in fact a
surprisingly simple close relationship between squeezing and
attainable entanglement. We have moreover quantified the maximal
degree of entanglement that can be achieved in a two-mode
subsystem, and we have identified the optimal entangling procedure
for the case of  two input modes. In view of recently proposed
applications of quantum information science, we hope that the
presented results as well as the employed techniques may prove
useful tools in the study of feasible sources of
continuous-variable entanglement.

We would like to thank S.\ Scheel and K.\ Audenaert for fruitful
discussions. This work has been supported by the ESF, the
A.v.-Humboldt Foundation, the DFG, and the European Union (EQUIP).

\end{multicols}


\begin{thebibliography}{99}
\bibitem{Theo}
    M.G.A.\ Paris, Phys.\ Rev.\ A {\bf 59}, 1615 (1999);
    S.\ Scheel, L.\ Kn{\"o}ll, T.\ Opatrny, and D.-G.\ Welsch,
    Phys.\ Rev.\ A {\bf 62}, 043803 (2000);
    M.S.\ Kim, W.\ Son, V.\ Bu\v zek, and P.L.\ Knight,
    Phys.\ Rev.\ A {\bf 65}, 032323 (2002);
    X.B.\ Wang, quant-ph/0204082, quant-ph/0204039;
    P. van Loock and S.L.\ Braunstein,
    Phys.\ Rev.\ Lett.\ {\bf 84}, 3482 (2000).
\bibitem{Exp}
    C.\ Silberhorn, P.K.\ Lam, O.\ Weiss,
    F.\ K{\"o}nig, N.\ Korolkova, G.\ Leuchs,
    Phys.\ Rev.\ Lett.\ {\bf 86}, 4267 (2001);
    N.\ Korolkova, C.\ Silberhorn, O.\ Gl{\"o}ckl,
    S.\ Lorenz, C.\ Marquardt, and
    G.\ Leuchs, Eur.\ Phys.\ D {\bf 18}, 229 (2002).
\bibitem{ReckZeil}
    M.\ Reck, A.\ Zeilinger, H.J.\ Bernstein, and P.\ Bertani,
        Phys.\ Rev.\ Lett.\ {\bf 73}, 58 (1994).
\bibitem{Neg}
    G.\ Vidal and R.F.\ Werner, Phys.\ Rev.\ A {\bf 65}, 032314 (2002).
\bibitem{SMD}
    R.\ Simon, N.\ Mukunda, and B.\ Dutta,
    Phys.\ Rev.\ A {\bf 49}, 1567 (1994).
\bibitem{CPM}
    B.\ Demoen, P.\ Vanheuverzwijn, and A.\
    Verbeure, Lett.\ Math.\ Phys.\ {\bf 2}, 161 (1977);
    J.\ Eisert and M.B.\ Plenio,
    Phys.\ Rev.\ Lett.\ {\bf 89}, 097901 (2002);
    J.\ Eisert, S.\ Scheel, and M.B.\ Plenio,
    Phys.\ Rev.\ Lett.\ {\bf 89}, 137903 (2002);
    J.\ Fiur{\'a}{\v s}ek, Phys.\ Rev.\ Lett.\ {\bf 89}, 137904 (2002);
    G.\ Giedke and J.I.\ Cirac, Phys.\ Rev.\ A {\bf 66}, 032316
    (2002).
\bibitem{BlochMessiah} C.\ Bloch and A.\ Messiah, Nuclear\ Physics
    {\bf 39}, 95 (1962).
\bibitem{linear}
        Here, {\it linearity} refers to the Hamiltonian,
    which is linear/non-linear w.r.t. creation and annihilation
    operators and {\it passive} means photon-number conserving.
\bibitem{WW}
    R.F.\ Werner and M.M.\ Wolf,
    Phys.\ Rev.\ Lett.\ {\bf 86}, 3658 (2001).

\bibitem{ABsplit} Note that the direct sum corresponds to a
split between modes rather than to a position/momentum split.
\bibitem{CiracSimon}
    R.\ Simon, Phys.\ Rev.\ Lett.\  {\bf 84}, 2726 (2000).
    L.-M.\ Duan, G.\ Giedke, J.I.\ Cirac, and P. Zoller,
    ibid.\  {\bf 84}, 2722 (2000).
\bibitem{distill}
    G.\ Giedke, L.-M.\ Duan, J.I.\ Cirac, and P.\ Zoller,
        Quant.\ Inf.\ Comp.\ {\bf 1}, 79 (2001).
\bibitem{det} Here we have used that for any symplectic matrix
$\det[{S}]=1$, and admissible covariance matrices satisfy
$\det[{\Gamma}]\geq 1$.
\end{thebibliography}
\end{document}